\documentclass{aa}
\usepackage{graphics}
\usepackage{tabularx}
\usepackage{amssymb}

\usepackage[authoryear]{natbib}
\bibpunct[]{(}{)}{;}{a}{}{,} 

\def\arcsec{\hbox{$^{\prime\prime}$}}

\begin{document}


\title{
Mid-infrared observations of the ultraluminous galaxies
\object{IRAS\,14348-1447}, \object{IRAS\,19254-7245},
and \object{IRAS\,23128-5919}\thanks{Based on observations with ISO, an
ESA project with instruments funded by ESA Member States (especially
the PI countries: France, Germany, the Netherlands and the United
Kingdom) and with participation of ISAS and NASA.}}

\author{V.~Charmandaris\inst{1} 
        \and
        O.~Laurent\inst{2,3}
        \and
        E.~Le Floc'h\inst{2}
        \and
        I.~F.~Mirabel\inst{2,4}
        \and
        M.~Sauvage\inst{2}
        \and
        S.~Madden\inst{2}
        \and
        P.~Gallais\inst{2}
        \and
        L.~Vigroux\inst{2}
        \and
        C.~J.~Cesarsky\inst{5}
       }

\institute{
Cornell University, Astronomy Department, Ithaca, NY 14853, USA
\and  
CEA/DSM/DAPNIA Service d'Astrophysique, F-91191 Gif-sur-Yvette, France  
\and
Max Planck Institut f\"ur extraterrestrische Physik, P.O. Box 1312, 85740  
Garching, Germany  
\and
Instituto de Astronom\'\i a y F\'\i sica del Espacio, Conicet, cc 67,
suc 28.  1428 Buenos Aires, Argentina
\and   
European Southern Observatory, Karl-Schwarzschild-Str, D-85748
Garching bei M\"unchen, Germany}

\offprints{Vassilis Charmandaris,
\email{vassilis@astro.cornell.edu}}
\titlerunning{ISOCAM observations of ultraluminous galaxies.}
\authorrunning{V. Charmandaris et al.}

\date{Received 1 March 2002 / Accepted 30 May 2002}


\abstract{ We present a study of the three ultraluminous infrared galaxies
IRAS\,14348-1447, IRAS\,19254-7245, and IRAS\,23128-5919, based on
mid-infrared (MIR) spectro-imaging (5--18\,$\mu$m) observations
performed with ISOCAM. We find that the MIR emission from each system,
which consists of a pair of interacting late type galaxies, is
principally confined to the nuclear regions with diameters of
1--2\,kpc and can account for more than 95\% of their IRAS 12\,$\mu$m
flux. In each interacting system, the galaxy hosting an active
galactic nucleus (AGN) dominates the total spectrum and shows stronger
dust continuum (12--16\,$\mu$m) relative to the Unidentified Infrared
Band (UIB) emission (6--9\,$\mu$m), suggestive of its enhanced
radiation field. The MIR dominant galaxy also exhibits elevated
15\,$\mu$m/H$\alpha$ and 15\,$\mu$m/K ratios which trace the high
extinction due to the large quantities of molecular gas and dust
present in its central regions.  Using only diagnostics based on our
mid-infrared spectra, we can establish that the Seyfert galaxy
IRAS\,19254-7245 exhibits MIR spectral features of an AGN while the
MIR spectrum of the Seyfert (or LINER) member of IRAS\,23128-5919 is
characteristic of dust emission principally heated by star forming
regions. 

}

\maketitle

\section{Introduction}

It is currently widely accepted that the majority of the most luminous
galaxies (L$_{bol}>10^{11}$\,L$_{\sun}$) in the local universe (z
$<0.3$) are luminous in the infrared, and include the ultraluminous
infrared galaxies (ULIRGs, L$_{\rm IR}>10^{12}$L$_{\sun}$) which emit
the bulk of their energy at infrared wavelengths \citep[ and
references therein]{Houck1984,Soifer1989,Sanders1996}. In those
systems most of the infrared emission seems to originate from their
dusty nuclear regions. Even though one of the principal heating
mechanisms for the lowest luminosity ($\lesssim 10^{11}$ L$_{\sun}$)
infrared galaxies is the stellar radiation field of young massive
stars, it is still unclear if the star formation is also the dominant
heating source for ULIRGs or whether one needs to invoke an active
galactic nucleus (AGN) and its strong radiation field as the central
engine responsible for the heating of the dust
\citep[see][]{Joseph99,Sanders99}. The presence of large quantities of
molecular gas has long been detected in the central regions of most
ULIRGs \citep[e.g.][]{Sanders1985,Sanders1991} leading to high
extinction of both their UV and optical radiation. As a result, since
it appears that most galaxies do harbor a super-massive, though often
quiescent, black hole \citep{Richstone1998}, one would expect to find
in their galactic nucleus observational evidence for a mixture of AGN
\citep{Sanders1988} and/or strong compact starburst regions
\citep{Condon1991} fueled by the high concentration of molecular gas
\citep{Bryant1999}. Observations in the mid-infrared (MIR), which are
less affected by absorption than shorter wavelengths
\citep[A$_{15\,\mu m}$ $\sim$\,A$_{V}$/70,][]{Mathis1990}, thus
provide a powerful probe of galactic central regions
\citep{Soifer2000,Soifer2001}.

As we discussed in \citet{Laurent2000}, the integrated MIR emission in
active galaxies is produced mainly by the interstellar dust which is heated
directly by the ionization field from young stars or an AGN. This is in
contrast to late type galaxies where the MIR (5--20$\mu$m) energy budget is
dominated by the reprocessed emission of star forming regions in their disk
and accounts for $\sim$\,15\% of their luminosity 
\citep{Dale2001,Helou2001,Roussel2001}. However, the main difficulty in
assessing the importance of the underlying physics in galactic nuclei, where
the spatial resolution is typically poor, is in separating the contribution
of star forming regions and the active nucleus from the integrated MIR
emission. The development, application, and general utility of MIR
diagnostics in nuclei of galaxies has already been demonstrated by 
\citet{Roche1991} and more recently by \citet{Genzel1998,Laurent2000}, as
well as by \citet{Dudley1999,Imanishi2000}. This was mainly
accomplished with the advent of ISOCAM and SWS on board ISO, with high
spatial and spectral resolution, as well as improved sensitivity in
the 3 to $\sim$40\,$ \mu$m wavelength range, thus allowing us to study
the nature of the heating sources in ULIRGs. More specifically it has
been shown by \citet{Lutz1998,Laurent1999b,Laurent2000,Tran2001} that
a nearby galaxy hosting a dominant AGN is clearly different in the MIR
from a starburst or a late type spiral. The most striking difference
is that the rather featureless MIR spectrum in AGN lacks the emission
bands at 6.2, 7.7, 8.6, 11.3 and 12.7\,$\mu$m, which are seen in late
type galaxies and are attributed to Polycyclic Aromatic Hydrocarbons
(PAHs) -- also often called Unidentified Infrared Bands (UIBs). One
may consider that this is simply due to the fact that its elevated MIR
continuum of the AGN overwhelms any UIB feature emission
\citep{Pier1992,Barvainis1987}. It seems inevitable that as the AGN
heats its dusty torus at T$\sim$\,1000\,K and the dust grains approach
sublimation temperatures, the more fragile molecules responsibly for
the UIB emission could be partly destroyed by a
photo-thermo-dissociation mechanism \citep{Leger1989}. Obviously this
picture is more complicated in distant galaxies since due to limited
spatial resolution the contribution of the star forming regions
surrounding an AGN would progressively enter into the beam and dilute
any AGN MIR signature
\citep[see][]{Laurent1999b}. When sufficient spatial resolution is available
to directly view the active nucleus, as is often the case in Seyfert 1
galaxies, the non-thermal emission from the AGN will dominate the spectrum.
Consequently, the spectrum can then be fitted by a power law and has a
``bump'' in the 4--5$\mu$m range. A 5--11\,$\mu$m study of a large sample of
Seyfert galaxies with ISO by \citet{Clavel2000} confirmed this picture,
concluding that Seyfert 2 galaxies have weaker MIR continuum. However, a
detailed analysis of the MIR spectra and images of the prototypical
Seyfert~2 galaxy NGC\,1068 by \citet{LeFloch2001} showed that if sufficient
spatial resolution is available and the AGN is extremely strong, even in the
case of a Seyfert 2 one can isolate the emission of the central engine from
the star forming regions which surround it. In that case the MIR spectrum of
the Seyfert 2 would also be a power law with the addition of a weak PAH
emission.

Despite this progress, several questions concerning the extent and spectral
characteristics of the MIR emission in active nuclei, as well as the
correlation between MIR and optical activity have not been fully examined.
Could broad band MIR photometry be used to probe the physical
characteristics of AGNs? In the present paper we try to address some of
these issues by studying the MIR spectral energy distribution (SED) of three
ultraluminous IRAS galaxies. Each IRAS source, the properties of which are
presented in Table~\ref{info}, consists of a merging pair of galaxies with
different levels of nuclear activity. The targets were specifically selected
as MIR bright and harboring an optically classified AGN. In section 2, we
describe the observations and in section 3 we present the details of our
study and analysis of the data for each system. A discussion followed by
concluding remarks is presented in section 4. Throughout this paper we
assume a Hubble constant H$_{0}$=75 km\,s$^{-1}$\,Mpc$^{-1}$ and q$_0$=1/2.


\section{Observations and data reduction}

Our MIR observations were obtained using ISOCAM, a 32$\times$32 pixel array 
\citep{CesarskyC1996} on board the ISO satellite \citep{Kessler1996}. Each
system was observed with broad band filters ranging from 5 to
18\,$\mu$m in a 2$\times$2 raster with 6 pixel offsets and a lens
producing a pixel field of view (PFOV) of 1.5{\arcsec}, resulting in a
final image of 57\arcsec$\times$57\arcsec. This enabled us to obtain
images with a spatial resolution of 3{\arcsec} (at 6\,$\mu$m) to
4.5{\arcsec} (15\,$\mu$m) limited by the pixel size at 6\,$\mu$m and
by the full width at half maximum (FWHM) of the point spread function
(PSF) at 15\,$\mu$m. We note the ISOCAM filters by their name and
central wavelength. The wavelength range in $\mu m$ covered by each
filter was: LW2 (5.0 -- 8.5), LW3 (12.0 -- 18.0), LW4 (5.5 -- 6.5),
LW6 (7.0 -- 8.5), LW7 (8.5 -- 10.7), LW8 (10.7 -- 12.0), LW9 (14.0 --
16.0). At subsequent sections in this paper we will refer to the
measured flux densities using the various filters as f$_{x\,\mu m}$
where \emph{x} is the central wavelength of each filter in microns.

Spectrophotometric observations were also obtained with the circular
variable filter (CVF) for IRAS 23128-5919, the brightest of our
sources. The CVF covers a spectral range from 5 to 16.5\,$\mu$m with a
1.5{\arcsec} PFOV and a spectral resolution of ~50. Each integration
step was composed of 12 images with 5.04 second integration time and
during the CVF scan the wavelength step varied between 0.05 and
1.11\,$\mu$m. Details on the observing parameters are summarized in
Table~\ref {param}.

The data were analyzed with the CAM Interactive Analysis software
(CIA\footnote{CIA is a joint development by the ESA astrophysics
division and the ISOCAM consortium}). A dark model taking into account
the observing time parameters was subtracted. Cosmic ray contamination
was removed by applying a wavelet transform method
\citep{Starck1997}. Corrections of detector memory effects were done
applying the Fouks-Schubert's method \citep{Coulais2000}. The flat
field correction was performed using the library of calibration data.
Finally, individual exposures were combined using shift techniques in
order to correct the effect of jittering due to the satellite motions
(amplitude $\sim$\,1\arcsec). A deconvolution using multiscale
resolution techniques \citep{Starck1999} was subsequently applied to
estimate the physical size of the quasi-point like sources responsible
for the infrared emission in our data (see section~\ref{sec:results}).

The details of the analysis of the ISOPHOT-S data of the three
galaxies, which we also include in this paper for reasons of
comparison, are published by \citet{Rigopoulou1999}.

Based on three different observations of IRAS 19254-7245 taken with
identical LW filters but with different roll angle, integration times
per exposure (2s and 5s) and PFOVs, as well as on similar analysis of
other ISOCAM-CVF and ISPHOT-S observations, we estimate that the
uncertainty of our photometry measurements is $\sim$\,20$\%$ (see
Table~\ref{phot}).

\begin{table*}[tbph]
\caption{Properties of the three IRAS systems.}
\label{info}
\begin{tabular}{ccccccccccc}
\hline\hline
Target & RA & DEC & z &\multicolumn{4}{c}{F$_{\nu}$\,(Jy)} & D$_{\rm L}$ &
log(L$_{\rm FIR}$) & log(L$_{\rm IR}$)\\ IRAS Name & J2000.0 & J2000.0 & & 
12\,$\mu$m & 25\,$\mu$m & 60\,$\mu$m & 100\,$\mu$m &
 (Mpc) & (L$_{\sun}$)          & (L$_{\sun}$)\\

19254-7245 & 19\fh31\fm21.6\fs & -72\fdg39\farcm20.8\farcs & 0.0617 & 0.22 & 1.24 & 
5.48 & 5.79 & 250 & 11.68 & 12.01\\

23128-5919 & 23\fh15\fm46.9\fs & -59\fdg03\farcm14.2\farcs & 0.0446 & 0.24 &
1.59 & 10.80 & 10.99 & 180 & 11.69 & 11.96\\

14348-1447 & 14\fh37\fm38.2\fs & -15\fdg00\farcm23.9\farcs & 0.0823  &
$\!\!\!\!\!<$0.14 & 0.49 & 6.87 & 7.07 & 335 & 12.05 & 12.27\\
\hline \hline
\end{tabular}
\newline
\par
\textbf{Table note:} The far-infrared and infrared luminosities are
calculated using L$_{\rm FIR}$=3.94$\times$10$^5\times$D(Mpc)$^2$(2.58$%
\times$\textit{f}$_{60}$+\textit{f}$_{100}$) and L$_{\rm IR}$=5.62$\times$10$^5\times$D(Mpc)$^2$(13.48$\times$\textit{f}$_{12}$+5.16$\times$
\textit{f}$_{25}$+2.58$\times$\textit{f}$_{60}$+\textit{f}$_{100}$)
respectively, where the luminosity distance is defined as D$_{\rm L}=
\frac{c}{H_0q_0^2}(zq_0+(q_0-1)(\sqrt{(1+2q_0z)}-1)$
\citep[see][]{Sanders1996}.
\end{table*}

\begin{table*}[tbph]
\caption{ ISOCAM observing parameters.}
\label{param}
\begin{tabular}{ccccccccccc}
\hline\hline
Target & ISOCAM Filter: & LW2 & LW3 & LW4 & LW6 & LW7 & LW8 & LW9 & CVF &
\\
& Filter Center: & 6.75\,$\mu$m & 15\,$\mu$m & 6\,$\mu$m & 7.75$\mu$m &
9.62\,$\mu$m & 11.4\,$\mu$m & 15\,$\mu$m & -- &  \\ \hline
IRAS 19254-7245$^1$ &  & 15.3 & 15.3 & 15.4 & 15.2 & 15.4 & 15.3 & 15.4 & --
&  \\
IRAS 19254-7245$^2$ &  & 7.1 & 7.0 & 11.3 & -- & 8.3 & -- & -- & -- &  \\
IRAS 19254-7245$^3$ &  & 3.4 & 3.6 & -- & -- & -- & -- & -- & -- &  \\
IRAS 23128-5919$^4$ &  & 7.2 & 7.0 & 11.5 & -- & 8.5 & -- & -- & -- &  \\
IRAS 23128-5919$^5$ &  & -- & -- & -- & -- & -- & -- & -- & 148.7 &  \\
IRAS 14348-1447$^6$ &  & 8.6 & 8.4 & -- & -- & -- & -- & -- & -- &  \\
\hline\hline
\end{tabular}
\par
\textbf{Table note:} The numbers following each galaxy denote the total
on-source exposure time (in minutes) for each filter used, and two galaxies
were observed more than once under different configurations, the details of
which are: (1) IRAS19254-7245 observed in proposal CAMACTI2 (PI I.F.
Mirabel), 7 LW filters, integration time per frame Tint=5s, pfov=1.5". (2)
IRAS19254-7245 observed in proposal CAMACTIV (PI I.F. Mirabel), 4 LW
filters, Tint=2s, pfov=1.5". (3) IRAS19254-7245 observed in proposal
SAM12N\_2 proposal (PI L. Spinoglio), 2 LW filters, Tint=2s, pfov=3". (4)
IRAS23128-5919 observed in proposal CAMACTIV (PI I.F. Mirabel), 4 LW
filters, Tint=2s, pfov=1.5". (5) IRAS23128-5919 observed in proposal
CAMACTI2 (PI I.F. Mirabel), CVF, Tint=5s, pfov=1.5". (6) IRAS14348-1447
observed in proposal CAMACTIV (PI I.F. Mirabel), 2 LW filters, Tint=2s,
pfov=1.5".
\end{table*}


\section{Results}

\subsection{Background and General Properties}

\label{sec:results}

The sensitivity and spatial resolution capabilities of ISOCAM enable
us to obtain deep maps of the MIR emission of each galaxy. Since the
interacting members of the IRAS galaxies are very close and are
point-like objects with one member typically dominating the MIR
emission, photometry measurements were treated with extra care. Our
approach was to fit the MIR PSF of the brightest component and to
subtract its contribution from the location of the neighboring,
fainter galaxy. We then performed aperture photometry on the fainter
component, using an aperture $\sim4.5''\times4.5''$. In spite of the
difference in their peak intensities, the relative positions of the
nuclei were very well known from deep near-IR imaging
\citep{Duc1997b}. Final aperture correction was applied to the flux of
each galaxy to account for the overall extension of the PSF. Our
measurements are presented in Table~\ref{phot}. We also include the
equivalent broad-band filter fluxes estimated from the ISOPHOT-S
spectra, which are found in good agreement with our data within the
photometric uncertainties. Since the galaxies were observed several
times under different ISOCAM configurations, more than one value is
often quoted for the same filter. This was done in order to display
the internal consistency of the different measurements and their
median value should be considered as the nominal flux density of each
galaxy.

ISOCAM has detected nearly $\sim$\,100\% of the 12\,$\mu$m IRAS flux
(see Table~\ref{info}) of these galaxies. Moreover, as it can be seen
from the images of the galaxies presented later in this section, no
extended extra-nuclear emission, has been detected in any of the
galaxies in the MIR. In all cases, the bulk of the flux coming from
these objects originates from a region less than 3--4.5{\arcsec} in
diameter (which corresponds to the FWHM of PSFs) associated with the
nuclei of the interacting galaxies.

\begin{table*}[htbp]
\caption{ISOCAM mid-infrared photometry of the sample.}
\label{phot}
\begin{tabular}{lrrrrrrrcc}
\hline \hline
Target             & \multicolumn{1}{c}{LW2}   & \multicolumn{1}{c}{LW3}   & 
                   \multicolumn{1}{c}{LW4}   & \multicolumn{1}{c}{LW6}   & 
                   \multicolumn{1}{c}{LW7}   & \multicolumn{1}{c}{LW8}   &
                   \multicolumn{1}{c}{LW9}   & \underline{LW3}           &
                   \underline{LW2}\\

IRAS              & \multicolumn{1}{c}{(mJy)} & \multicolumn{1}{c}{(mJy)} & 
                   \multicolumn{1}{c}{(mJy)} & \multicolumn{1}{c}{(mJy)} & 
                   \multicolumn{1}{c}{(mJy)} & \multicolumn{1}{c}{(mJy)} & 
                   \multicolumn{1}{c}{(mJy)} & LW2                       &
                   LW4\\ \hline

19254S$^{1}$     & 106.9$\pm$10.7            & 284.0$\pm$28.4            &
                    90.0$\pm$9.1             & 150.1$\pm$15.0            &
                    91.2$\pm$9.1             & 107.5$\pm$10.8            &
                   337.5$\pm$33.8            &   2.7$\pm$0.4             &
                     1.2$\pm$0.2\\ 

19254S$^{2}$     & 103.6$\pm$11.0            & 278.9$\pm$29.1            &
                    87.3$\pm$11.2            & \multicolumn{1}{c}{--}    & 
                    97.1$\pm$10.5            & \multicolumn{1}{c}{--}    & 
                   \multicolumn{1}{c}{--}    &   2.7$\pm$0.4             &
                    1.2$\pm$0.2\\

\\
19254N$^{1}$     &   4.8$\pm$0.5             &   5.9$\pm$0.7             &
                     1.9$\pm$0.4             &   8.3$\pm$0.9             &
                     5.1$\pm$0.6             &   5.9$\pm$0.7             &
                     5.4$\pm$1.0             &   1.2$\pm$0.2             &
                     2.5$\pm$0.6\\

19254N$^{2}$     &   3.1$\pm$1.0             &   7.5$\pm$2.4             &
                     1.5$\pm$2.6             & \multicolumn{1}{c}{--}    & 
                     4.5$\pm$1.7             & \multicolumn{1}{c}{--}    & 
                   \multicolumn{1}{c}{--}    &   2.4$\pm$1.1             &
                     2.1$\pm$3.6\\

\\
19254$^{1}$      & 111.7$\pm$11.2            & 289.9$\pm$29.0            &
                    91.9$\pm$9.2             & 158.4$\pm$15.9            &
                    96.3$\pm$9.6             & 113.4$\pm$11.3            &
                   342.9$\pm$34.3            &   2.6$\pm$0.4             &
                     1.2$\pm$0.2\\

19254$^{2}$      & 106.7$\pm$11.0            & 286.4$\pm$29.2            & 
                    88.8$\pm$11.5            & \multicolumn{1}{c}{--}    & 
                    97.1$\pm$10.6            & \multicolumn{1}{c}{--}    & 
                   \multicolumn{1}{c}{--}    &   2.7$\pm$0.4             &
                     1.2$\pm$0.2\\

19254$^{3}$      & 114.8$\pm$12.5            & 290.2$\pm$11.5            &
                   \multicolumn{1}{c}{--}    & \multicolumn{1}{c}{--}    & 
                   \multicolumn{1}{c}{--}    & \multicolumn{1}{c}{--}    & 
                   \multicolumn{1}{c}{--}    &   2.5$\pm$0.4             &
                   \multicolumn{1}{c}{--}\\

19254$^{\dag}$   & 113.0$\pm$2.8             & \multicolumn{1}{c}{--}    &
                    85.6$\pm$3.7             & 135.7$\pm$4.1             &
                   110.4$\pm$4.6             & 116.9$\pm$13.9            &
                   \multicolumn{1}{c}{--}    & \multicolumn{1}{c}{--}    &
                     1.3$\pm$0.1\\
\hline

23128S$^{4}$     &  70.8$\pm$7.1             & 228.3$\pm$22.9            &
                    48.5$\pm$5.0             & \multicolumn{1}{c}{--}    &
                    67.3$\pm$6.8             & \multicolumn{1}{c}{--}    &
                   \multicolumn{1}{c}{--}    &   3.2$\pm$0.5             &
                     1.5$\pm$0.2\\ 

23128S$^{5}$     &  77.5$\pm$1.6             & 262.3$\pm$4.3             &
                    50.6$\pm$2.3             & 116.0$\pm$2.6             &
                    79.9$\pm$2.0             & 106.2$\pm$3.6             &
                   277.1$\pm$6.6             &   3.4$\pm$0.1             &
                     1.5$\pm$0.1\\ 
\\

23128N$^{4}$     &  38.6$\pm$3.9             &  88.5$\pm$9.1             &
                    19.8$\pm$2.2             & \multicolumn{1}{c}{--}    &
                    26.7$\pm$2.8             & \multicolumn{1}{c}{--}    &
                   \multicolumn{1}{c}{--}    &   2.3$\pm$0.3             &
                     2.0$\pm$0.3\\

23128N$^{5}$     &  34.8$\pm$1.2             &  90.4$\pm$2.1             &
                    19.9$\pm$2.0             &  53.4$\pm$1.7             &
                    34.7$\pm$1.5             &  51.3$\pm$2.4             &
                    90.9$\pm$3.2             &   2.6$\pm$0.1             &
                     1.7$\pm$0.2\\
\\

23128$^{4}$      &  109.4$\pm$11.0           & 316.8$\pm$31.7            &
                     68.3$\pm$7.0            & \multicolumn{1}{c}{--}    &
                     94.0$\pm$9.5            & \multicolumn{1}{c}{--}    &
                   \multicolumn{1}{c}{--}    &   2.9$\pm$0.4             &
                      1.6$\pm$0.2 \\

23128$^{5}$      &  112.3$\pm$2.6            & 352.7$\pm$7.2             &
                     70.5$\pm$3.4            & 169.4$\pm$4.4             &
                    114.5$\pm$3.1            & 157.5$\pm$6.1             &
                    368.0$\pm$10.7           &   3.1$\pm$0.1             &
                      1.6$\pm$0.1 \\

23128$^{\dag}$   &  123.6$\pm$3.2            & \multicolumn{1}{c}{--}    &
                     70.0$\pm$3.6            & 163.0$\pm$4.9             &
                    134.5$\pm$4.8            & 140.4$\pm$14.2            &
                   \multicolumn{1}{c}{--}    & \multicolumn{1}{c}{--}    &
                      1.8$\pm$0.1 \\
\hline

14348S$^{6}$     &  21.8$\pm$4.6             &  73.9$\pm$9.0             &
                   \multicolumn{1}{c}{--}    & \multicolumn{1}{c}{--}    &
                   \multicolumn{1}{c}{--}    & \multicolumn{1}{c}{--}    &
                   \multicolumn{1}{c}{--}    &   3.4$\pm$0.8             &
                   \multicolumn{1}{c}{--}\\ 
\\
14348N$^{6}$     &  11.3$\pm$3.6             &  22.9$\pm$5.0             &
                   \multicolumn{1}{c}{--}    & \multicolumn{1}{c}{--}    &
                   \multicolumn{1}{c}{--}    & \multicolumn{1}{c}{--}    &
                   \multicolumn{1}{c}{--}    &   2.0$\pm$0.8             &
                   \multicolumn{1}{c}{--}\\
\\
14348$^{6}$      &  33.1$\pm$5.8             &  96.8$\pm$10.3            &
                   \multicolumn{1}{c}{--}    & \multicolumn{1}{c}{--}    &
                   \multicolumn{1}{c}{--}    & \multicolumn{1}{c}{--}    &
                   \multicolumn{1}{c}{--}    &   2.9$\pm$0.6             &
                   \multicolumn{1}{c}{--}\\

14348$^{\dag}$   &  37.9$\pm$1.6             & \multicolumn{1}{c}{--}    &
                    16.9$\pm$2.0             & 52.6$\pm$2.4              &
                    34.4$\pm$3.3             & 14.1$\pm$10.7             &
                   \multicolumn{1}{c}{--}    & \multicolumn{1}{c}{--}    &
                     2.2$\pm$0.3\\

\hline
\hline
\end{tabular}
\textbf{Table note:} For each interacting system, we have measured the
integrated flux of the individual galaxies resolved by ISOCAM and
marked the southern and the northern galaxies with (S) and (N)
respectively. We used the same notations as in Table 2 for identifying
the different sets of ISOCAM observations, labeled (1) through (6). As
all three galaxies were also observed with ISOPHOT-S and one with the
CVF, we also provide the equivalent broad-band filter flux estimates
(using the known filter band-passes and transmission curves) based on
those spectra marked with a $\dag$ for ISOPHOT-S and a (5) for the
CVF. The errors given for each measurement are statistical derived by
adding the 1$\sigma$ rms map to the systematic error of 10\% commonly
associated with the transient correction. Absolute flux uncertainties
are estimated to be $\pm$ 20$\%$. For the cases where we present
multiple measurements for a target their median value should be
considered as its nominal flux density.
\end{table*}

As it has been discussed in several papers describing ISO observations
\citep[i.e.][ and references therein]{Laurent2000} the MIR emission of
spiral galaxies observed by ISOCAM originates from a number of
physical processes, with two dust heating mechanisms typically
prevailing. One is the thermal emission produced by
thermally-fluctuating, small grains ($\sim$10\,nm) heated by the
interstellar radiation field, observed between 12\,$\mu$m and
18\,$\mu$m in areas of strong radiation environments and is often
sampled by the LW3 filter. The second is due to the UIBs, which
originated from complex 2-dimensional aromatic molecules having C=C
and C\---H bonds and can be seen at 6.2, 7.7, 8.6, 11.3 and
12.7\,$\mu$m in the ISOCAM wavelength range. The emission in these
bands can be observed either with the CVF or using a sequence of
narrow-band filters. An absorption feature due to silicates is often
observed at 9.7\,$\mu$m and can be measured using the LW7
(8.5-10.7\,$\mu$m) filter.  Finally, two forbidden emission lines due
to [NeII] at 12.8\,$\mu$m and [NeIII] at 15.5\,$\mu$m can be detected
in the CVF mode. A contribution to the MIR spectrum by a third
component, the Rayleigh-Jeans tail of an old stellar population, is
generally negligible in late type galaxies where the hot dust emission
dominates. This MIR emission directly arising from stellar photosphere
is detected in early type galaxies
\citep{Madden1997}.

Analysis of a wealth of ISOCAM data has shown that the flux ratio of
the broad band filters centered at 15\,$\mu$m and 6.75\,$\mu$m
(LW3/LW2 or f$_{15\mu m}$/f$_{6.7\mu m}$) provides a diagnostic of the
dominant global MIR emission characteristic of \ion{H}{ii} regions,
the diffuse interstellar medium or photo-dissociation regions
\citep{Verstraete1996,CesarskyD1996b, Dale2001, Roussel2001}. It has
been shown that while quiescent star forming regions typically have
f$_{15\mu m}$/f$_{6.7\mu m}$$\sim$\,1, in active sites of massive star
formation this ratio increases due to the increasing contribution of
the continuum emission in the 15$\mu$m bandpass \citep{Sauvage1996,
Mirabel1998,Vigroux1999,Dale2001}. However, one should note that the
use of this indicator alone is not sufficient to distinguish between
the MIR spectrum due to star formation or an AGN, since in AGNs the
hot dust continuum arising from the torus also has f$_{15\mu
m}$/f$_{6.7\mu m}$ $>$1. Such a degeneracy may be resolved using the
flux ratio of the 6.75\,$\mu$m LW2 filter (sampling the 6.2 and
7.7\,$\mu$m UIBs) to the narrower LW4 filter which is centered at
6.0\,$\mu$m only contains the 6.2\,$\mu$m UIB.  As the continuum
variation between these two filters is negligible, the f$_{6.7\mu
m}$/f$_{6\mu m}$ (LW2/LW4) ratio estimates the intensity of UIBs
relative to the underlying continuum \citep[see Fig.5 of
][]{Laurent2000}. The closer f$_{6.7\mu m}$/f$_{6\mu m}$ is to 1, the
stronger the continuum is. Since AGNs have weaker UIBs than
starbursts, \citet{Laurent2000} proposed to use the combination of the
f$_{15\mu m}$/f$_{6.7\mu m}$ and f$_{6.7\mu m}$/f$_{6\mu m}$ colors to
differentiate between the two mechanisms contributing to the MIR
emission. Clearly there is a redshift dependence of this diagnostic
due to the K-correction of the SEDs, but since the redshifts of our
targets are small, these indicators can be applied
\citep{Laurent1999a}.

Using a large sample of galaxies in the Virgo cluster
\citet{Boselli1997} studied the properties of their MIR emission,
normalized to the mass of these galaxies. This was done by examining
the ratios of the f$_{6.7\mu m}$ (LW2) and f$_{15\mu m}$ (LW3) flux
densities to the K band light, which scales with stellar mass of the
galaxy, and it was found that the typical f$_{15\mu m}$/K ratio for a
late type spiral ranges between 1 and 10. In Table~\ref{k_ha}, we
present those ratios for our sample and we find that even though their
active nuclei must contribute some non-thermal emission in the K band
the ratios are considerably larger. This can be attributed to a
combination of increased thermal dust emission along with a wavelength
dependent absorption, which, in highly obscured sources, may decrease
their K band flux. Such an example is Arp\,220 which displays a ratios
f$_{15\mu m}$/K$\sim $\,30 \citep{Charmandaris2002}. Two more ratios
of the LW3 and LW2 over the H$\alpha$ line flux density are also
included in Table~\ref{k_ha} for reasons of completeness. It has been
established that in normal spirals, both filters mostly trace the MIR
flux arising from the reprocessing of ionising radiation which is
observed in the optical via the H$\alpha$ line
\citep{Sauvage1996,Roussel2001,Dale2001}. Since in more active
galaxies, the H$\alpha$ emission is strongly affected by the
absorption, these ratios could be used to quantify the level of
absorption\footnote{For a typical visual absorption range of 1-3\,mag,
f$_{15\mu m}$/H$\alpha$ varies between 10 and 80
\citep[see][]{Sauvage1996,Roussel2001}} even though one should be
cautious in their quantitative interpretation since the ratios may
saturate toward extreme starbursts \citep{Roussel2001}. We present the
LW2/H$\alpha$ mainly for comparison, as the most interesting indicator
is clearly the one involving the LW3 filter which directly traces the
continuum of hot dust emission emitted by the small grains.

\begin{table}[!htb]
\caption{Broad band color ratios. H$\alpha$ and K band fluxes are
from \citet{Duc1997a} except the H$\alpha$ flux of IRAS 14348-1447
\citep[see][]{Veilleux1995}. LW2, LW3 and K are in mJy and H$\alpha$
in $10^{-13}$erg cm$^{-2}$s$^{-1}$.}
\label{k_ha}
\begin{tabular}{lcccc}\hline \hline
Target          &   \underline{LW2} &  \underline{LW2} & \underline{LW3} 
                &  \underline{LW3}\\
       &   H$\alpha$ &  K  &   H$\alpha$      &  K\\ \hline

IRAS 19254-7245\,(S)& 91.3 & 12.4 & 224.5 & 30.4 \\ 
IRAS 19254-7245\,(N)& 78.4 & 1.2  & 85.4  & 1.3  \\ 
\\
IRAS 23128-5919\,(S)& 63.5  & 9.9 & 283.8 & 44.3 \\
IRAS 23128-5919\,(N)& 31.2  & 8.2 & 85.2  & 22.5 \\
\\
IRAS 14348-1447\,(S)& 125.7 & 8.1 & 332.9 & 21.5 \\ 
IRAS 14348-1447\,(N)& 161.2 & 6.3 & 263.3 & 10.3 \\ 
\hline \hline
\end{tabular}
\end{table}

Finally, in Table~\ref{global} we also present the MIR luminosities of
both the LW2 and LW3 filters for each galaxy of our sample. One can
clearly see that despite the activity in these systems, the MIR
spectrum contains only a small fraction ($<$5\%) of their energy which
is mostly emitted at longer wavelengths in the far-infrared (FIR).
This is in sharp contrast from what is seen in normal late type
galaxies where $\sim$15\% of the luminosity is emitted between
5--20$\mu$m \citep{Dale2001}. In the same table we include the L$_{\rm
IR}$(L$_{\sun}$)/M$_{\rm H_2} $(M$_{\sun}$) ratio which traces the
efficiency of molecular gas consumption, via either star formation or
AGN activity, as well as the production of high energy photons which
in-turn are reprocessed into infrared via dust absorption and/or
scattering. As expected the reported values for our sample are typical
of ultraluminous galaxies while ormal spiral galaxies such as the
Milky Way have a ratio of 1--10\,L$_{\sun}$\,M$_{\sun}$$^{-1}$, while
starbursts such as M82 display higher
$\sim$100\,L$_{\sun}$\,M$_{\sun}$$^{-1}$ values
\citep[see][]{Sanders1986,Wild1992}.

\begin{table}[!htb]
\caption{ Global characteristics. The IR luminosities and the H$_2$ mass are
in solar units \citep[see][]{Mirabel1990}.}
\label{global}
\begin{tabular}{lccccc}\hline \hline
Target                  &   L$_{\rm LW2}$    &  L$_{\rm LW3}$  & \underline{L$_{\rm LW2}$} 
& \underline{L$_{\rm LW3}$} & \underline{L$_{\rm IR}$}\\

IRAS name               &  (10$^9$L$_{\sun}$) &(10$^9$L$_{\sun}$)& L$_{\rm IR}$             
& L$_{\rm IR}$             & M$_{\rm H_2}$\\ \hline

19254-7245\,(S)         &   53.8         & 42.9     & -- & -- & --\\
19254-7245\,(N)         &    1.1         &  0.9     & -- & -- & --\\
19254-7245              &   54.9         & 43.8     & 0.05 & 0.04   & 34.1 \\
\\                                                         
23128-5919\,(S)         &   17.7         & 19.2     & -- & -- & --\\
23128-5919\,(N)         &    9.6         &  7.5     & -- & -- & --\\
23128-5919              &   27.3         & 26.7     & 0.03 & 0.03 & 70.2\\
\\                                                         
14348-1447\,(S)         &   18.9         & 21.6     & -- & -- & --\\ 
14348-1447\,(N)         &    9.8         &  6.7     & -- & -- & --\\ 
14348-1447              &   28.7         & 28.3     & 0.02 & 0.02 & 31.0\\ 
\hline \hline
\end{tabular}
\end{table}


Let us now review the MIR properties of each system in detail.

\subsection{IRAS\,19254-7245}

\begin{figure*}[!ht]
\vspace{5mm} \hspace{-3.5mm} \resizebox{\hsize}{!}{%
\includegraphics{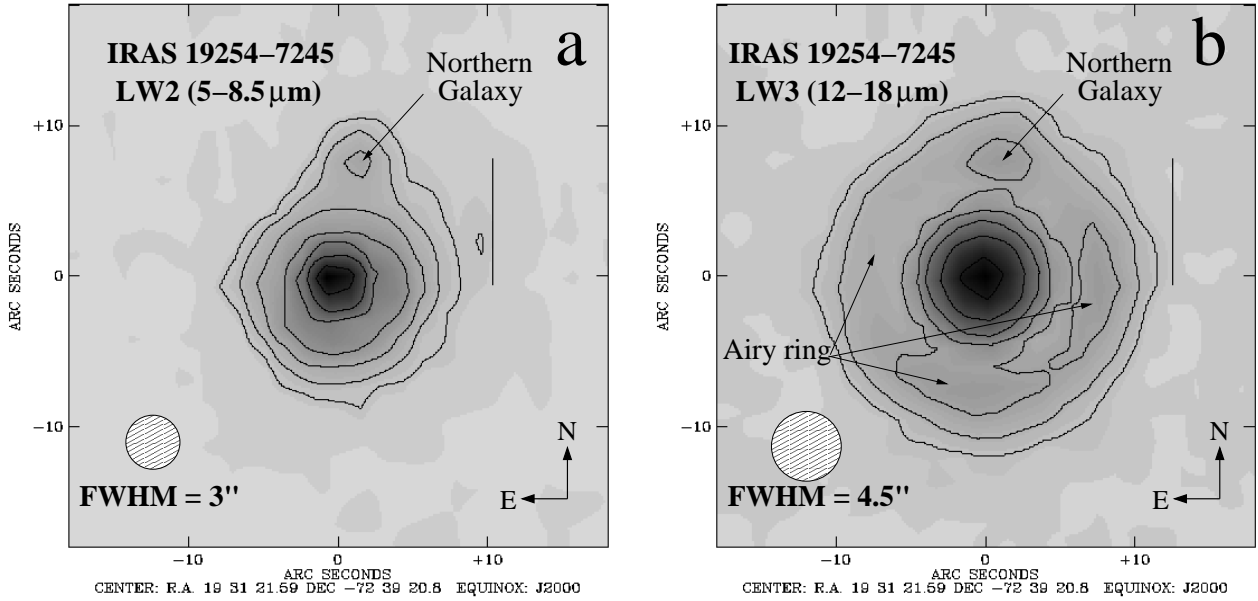}}
\caption{ a) Image of IRAS\,19254-7245 observed with ISOCAM at 6.75\,$%
\protect\mu$m (LW2 filter). The contours are 5, 10, 20, 40, 80, 160 and 320\,%
$\protect\sigma$ ($\protect\sigma$=0.027\,mJy\,pixel$^{-1}$). b) Image of
IRAS\,19254-7245 observed with ISOCAM at 15\,$\protect\mu$m (LW3 filter).
The contours are 5, 10, 20, 40, 80, 160 and 320\,$\protect\sigma$ ($\protect%
\sigma$=0.041\,mJy\,pixel$^{-1}$). Note that the PSF has a clear asymmetry
in the southwest direction of the Airy ring. The two vertical bars
correspond to 10\,kpc. }
\label{ir19254_im}
\end{figure*}

The ultraluminous infrared galaxy IRAS\,19254-7245, also known as the
``Superantennae'' is the result of a collision between two gas-rich
spiral galaxies separated by 10 kpc (8.5\arcsec) in projection and
displays extremely long tidal tails extending to 350\,kpc
\citep{Mirabel1990}. Only the MIR emission originating from the
nuclear regions of the galaxies is detected in our images
(Figure~\ref{ir19254_im}), and there is no evidence for emission
extending toward the direction of the tails. Even the northern nucleus
is marginally above the sensitivity limit $\sim$\,1\,mJy at
3\,$\sigma$ (see Table~\ref{phot}).

Using optical spectroscopy, the southern galaxy has been classified as
a Seyfert 2 with an observed FWHM of $\sim$1700\,km\,s$^{-1}$ in both
permitted and forbidden lines \citep{Mirabel1990,Duc1997a}. The
presence of an active nucleus is further suggested by the IRAS
criteria for selecting Seyferts, since the ratio of its 25\,$\mu$m to
the 60\,$\mu$m IRAS flux density is greater than 0.2
\citep[see][]{deGrijp1985}, while its optical and near-infrared colors
indicate a strong contribution from a non-thermal component, likely
originating from an AGN, as well as emission from very hot dust
($\sim$\,1000\,K) \citep{Vanzi2002}. Evidence of massive star
formation is also seen in the nuclear regions as emission line
splitting which has been attributed to a biconical outflow
\citep{Colina1991}. The kinetic energy necessary for this to occur can
only be produced by supernova explosions or stellar winds further
suggesting high star formation rates
\citep[150\,M$_{\sun}$\,yr$^{-1}$,][]{Colina1991}. Ground-based MIR
observations at 10\,$\mu$m show that more than 80\,$\%$ of the total
flux originates from the Seyfert 2 (the southern galaxy).  The
spectrum of the northern galaxy has much weaker emission lines in
H$\alpha$ and [NII], typical of a starburst or LINER
\citep{Colina1991}. More recently HST imaging provided new evidence
that a double nucleus may be present in both the north and southern
components of the Superantennae \citep{Borne1999}, suggesting a
multiple merger origin of the system.

\begin{figure*}[!ht]
\vspace{5mm} \hspace{-3.5mm} \resizebox{\hsize}{!}{\includegraphics{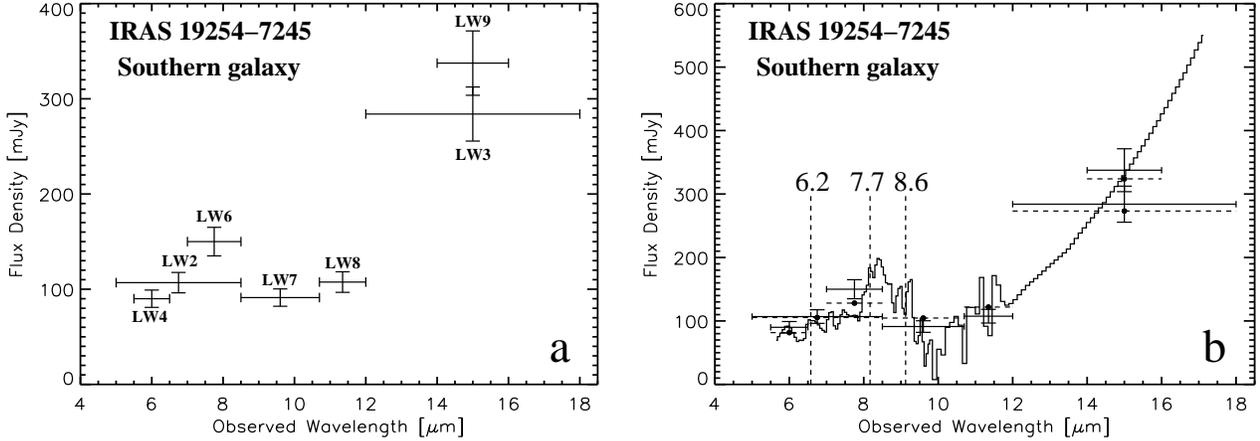}}
\caption{ a) The MIR spectral energy distribution (SED) of the southern
component of IRAS\,19254-7245 based on the 7 ISOCAM broad band
filters. The width of each filter is indicated by a horizontal line
and the photometric uncertainties by the vertical lines. b) Same as a)
including the ISOPHOT spectrum of the southern component. Since the
ISOPHOT beam covered the both galaxies the latter was estimated after
subtracting the template MIR spectrum of M82 for the emission from the
northern companion which is, as we see as well in
Figure~\ref{ir19254n_spec}, necessary to explain the high LW9/LW3
ratio. The ISOPHOT data end at $\sim$12$\protect\mu m$, but for
instructive purposes we mark the continuum between 12 and
17\,$\protect\mu$m with a power law after having subtracted the weak
contribution from the northern galaxy. To visually estimate the
quality of the fit we include again for comparison the observed flux
in this galaxy presented in the left panel. The positions of the main
UIB features, redshifted due to the distance of the galaxy are also
marked. The elevated MIR fluxes near 5$\mu$m suggest that this galaxy,
classified optically as Seyfert 2, has the typical MIR characteristics
of an AGN.}
\label{ir19254s_spec}
\end{figure*}

\begin{figure*}[!ht]
\vspace{5mm} \hspace{-3.5mm} \resizebox{\hsize}{!}{\includegraphics{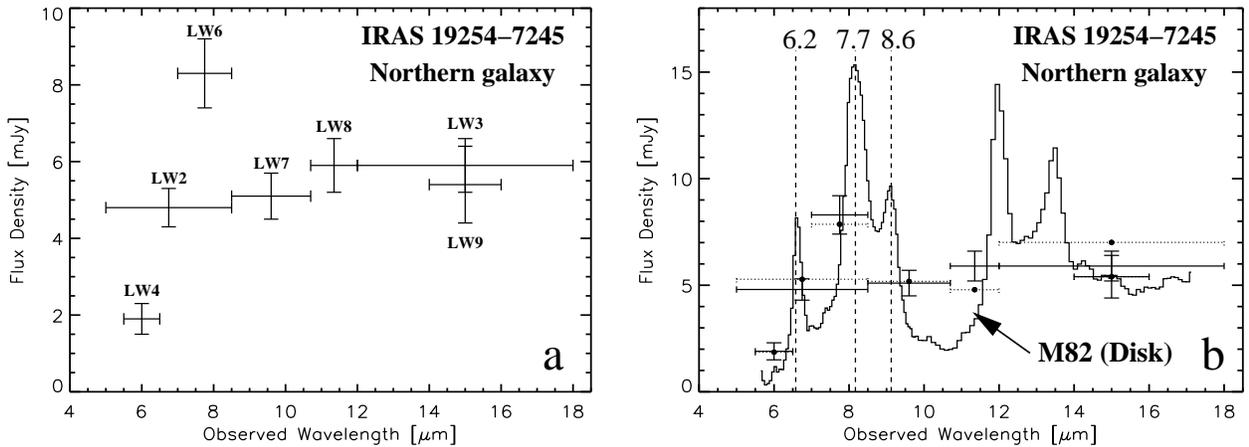}}
\caption{ 
a) The MIR spectral energy distribution (SED) of the northern
component of IRAS\,19254-7245 based on the 7 ISOCAM broad band
filters. Again the width of each filter is indicated by a horizontal
line and the photometric uncertainties by the vertical lines. b) The
same SED including a template fit of CVF spectrum (solid line) from a
quiescent star forming region in the disk of M82. The spectrum of M82
has been normalized to the flux density sampled by the LW9 filter of
IRAS\,19254-7245, which traces essentially the VSG component. The
dotted lines are the equivalent broad-band filter fluxes of the M82
template. The small offset between those and the actual measurements
of IRAS\,19254-7245, indicate that the northern galaxy is dominated in
the MIR by starburst activity.} \label{ir19254n_spec}
\end{figure*}

\begin{figure}[!ht]
\vspace{5mm} \hspace{-3.5mm} \resizebox{\hsize}{!}{\includegraphics{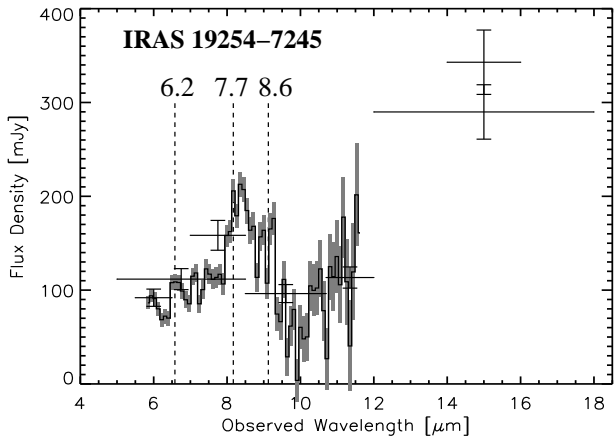}}
\caption{The integrated MIR SED of IRAS\,19254-7245. Our measurements using
the 7 ISOCAM broad band filters (solid horizontal lines) are compared
to the ISOPHOT-S spectrum. The photometric uncertainties of the
ISOPHOT-S are indicated by the hashed zones. The positions of the main
UIB features, redshifted due to the distance of the galaxy are again
marked.}
\label{ir19254s_phot}
\end{figure}

Based on the photometry of Table~\ref{phot}, we present in
Figures~\ref {ir19254s_spec} and \ref{ir19254n_spec} the MIR spectral
energy distribution for each galaxy, while the integrated MIR SED of
the whole system is shown in Figure~\ref{ir19254s_phot}. In the latter
we also compare our data with the spectrum obtained with ISOPHOT-S,
the beam of which spatially covered the full emission of
IRAS\,19254-7245. The extreme difference in the MIR intensities
between the southern and northern members are apparant as well as the
constrasts in their spectral shape.

More than 95 $\%$ of the MIR emission of IRAS\,19254-7245 originates
from the southern Seyfert 2 galaxy which displays a peculiar spectrum
with a dominant thermal emission at 15\,$\mu$m (f$_{15\mu
m}$/f$_{6.7\mu m}$ $\sim$\,2.7) and weak UIBs (f$_{6.7\mu m}$/f$_{6\mu
m}$$\sim$\,1.2). This strong continuum relative to the UIB emission
can be the consequence of a high radiation field density mainly
produced in ionized regions close to young stars \citep{Mirabel1998}
or AGN \citep{Laurent2000}. On the contrary, the northern galaxy has
strong UIBs (f$_{6.7\mu m}$/f$_{6\mu m}$$\sim$\,2.5) and faint thermal
emission at 15\,$\mu$m (f$_{15\mu m}$/f$_{6.7\mu m}$$\sim$\,1.2),
which is typical of MIR emission from normal spiral galaxies with cool
IRAS colors \citep{Dale2001,Roussel2001}. Comparison of in its
broad-band SED with the template SED of a quiescent star forming
region within the disk of M82 \citep{Laurent2000} illustrates that
they are in a fairly good agreement (Figure~\ref{ir19254n_spec}).

The absolute luminosities presented in Table~\ref{global} shows that
the MIR emission originating from the southern Seyfert 2 galaxy is by
far the strongest in our sample although the most luminous FIR source
is IRAS\,14348-1447 (see Table~\ref{info}). One may also note that the
flux density near 5\,$\mu$m does not reach zero level but is
$\sim$100\,mJy, suggesting the presence of a hot dust component, which
as discussed in the previous section is a clear sign of a hot dusty
torus of an AGN \citep{Laurent2000}. Similarly, one can draw the same
conclusion by observing the combination of the f$_{15\mu
m}$/f$_{6.7\mu m}$ and f$_{6.7\mu m}$/f$_{6\mu m}$ flux ratios. In
IRAS\,19254-7245S, the low f$_{6.7\mu m}$/f$_{6\mu m}$ indicates weak
UIB emission while f$_{15\mu m}$/f$_{6.7\mu m}$$\sim$\,2.7, a value
somewhat lower than other well studied starburst galaxies such as
Arp\,220 \citep[f$_{15\mu m}$/f$_{6.7\mu
m}$$\sim$\,3.9,][]{Charmandaris1999b} or the extremely strong
starburst region in the Cartwheel \citep[f$_{15\mu m}$/f$_{6.7\mu
m}$$\sim$\,5.2,][]{Charmandaris1999a}. This effect can be understood
since the hot continuum produced by an AGN at short MIR wavelengths
would cause the flux in the 6--10\,$\mu$m range to increase more
relative to the increase observed between 12--16\,$\mu$m and as result
it would be added the UIB emission sampled by the LW2 filter.

Could the large difference in the MIR brightness between the north and
south component in IRAS\,19254-7245 be related to the additional
contribution of the AGN? Studies of the dynamical evolution of this
system suggest that the starburst time scale is much shorter than the
dynamical age of the merger \citep{Mihos1998}. Even though we can not
quantify accurately the fraction of MIR luminosity due to the AGN
activity, it appears that the southern component of IRAS\,19254-7245
has reached an AGN dominant phase, however short this may be, after an
initial phase of strong starburst activity (see \citet{Laurent2000}
and \citet{Genzel1998} for details on the MIR AGN/starburst fraction
of this and other galaxies). The MIR properties of the northern
nucleus are similar to a normal spiral galaxy which indicates that
even if a starburst did occur in it at some point, it has by now
subsided and the star formation is progressing in a more quiescent
rate.

Finally, we note that the southern galaxy exhibits higher f$_{15\mu
m}$/H$\alpha$ ($\sim$\,225) compared to that inthe north
($\sim$\,85). We interpret this effect as a consequence of higher dust
concentration and stronger absorption in the southern nucleus since
near AGNs high column densitis of molecular gas are typically
observed. The southern galaxy also has a higher f$_{15\mu m}$/K ratio
than that in the north, which has an f$_{15\mu m}$/K ratio of a normal
spiral galaxy,consistent with its overall MIR spectral features
(Table~\ref{k_ha}).


\subsection{IRAS\,23128-5919}

\begin{figure*}[!ht]
\vspace{0mm} \hspace{0mm} \resizebox{\hsize}{!}{\includegraphics{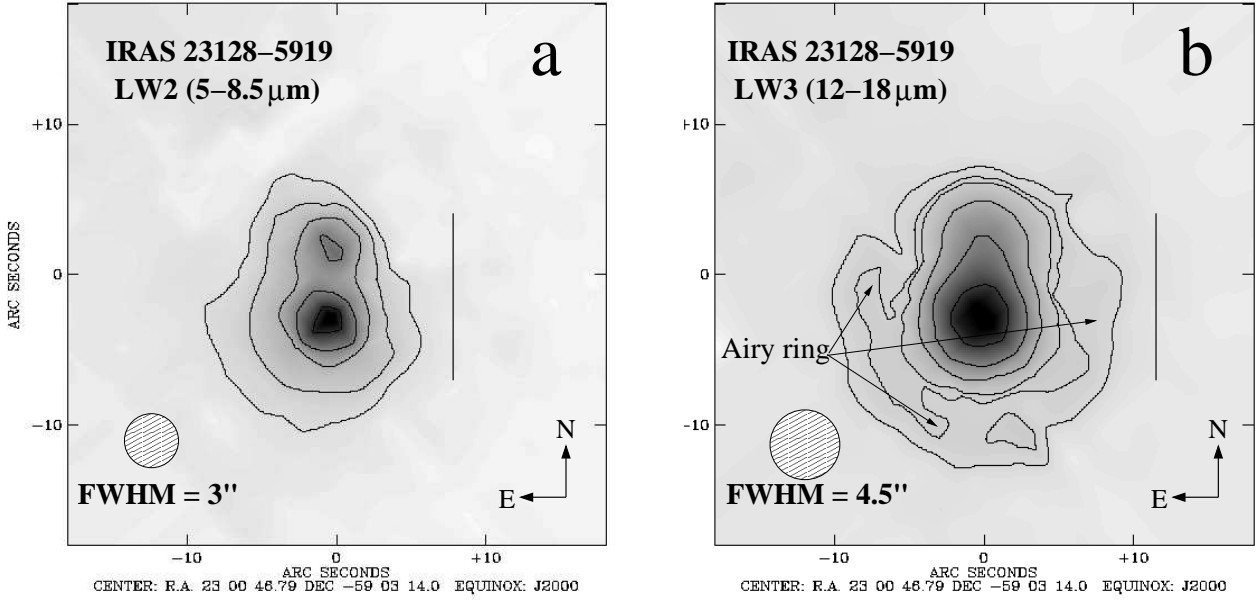}}
\caption{ 
a) Image of IRAS\,23128-5919 observed with ISOCAM at 6.75\,$\mu$m (LW2
filter). The contours are 5, 10, 20, 40 and 80\,$\sigma$
($\sigma$=0.055\,mJy\,pixel$^{-1}$). b) Image of IRAS\,23128-5919
observed with ISOCAM at 15\,$\mu$m (LW3 filter).  The contours are 5,
8, 10, 20, 40 and 80\,$\sigma$ ($\sigma$
=0.093\,mJy\,pixel$^{-1}$). Note the asymmetry of the PSF towards the
southeast of the Airy ring. The two vertical bars correspond to
10\,kpc.}
\label{ir23128_im}
\end{figure*}

This system consists of two merging galaxies in a rather late stage of
their interaction, the nuclei of which are separated by a projected
distance of 4 kpc (5\arcsec) (Figure~\ref{ir23128_im}). Two tidal
tails 40\,kpc stretch in opposite directions
\citep{Bergvall1985,Mihos1998}.

Based on optical studies, the northern galaxy is classified as a
starburst, while it is unclear whether the southern one is a Seyfert,
a starburst or a LINER \citep{Duc1997a}.  Optical spectroscopy of the
southern nucleus shows a relatively high ionization state having
emission lines with wings of $\sim$ 1500\,km\,s$^{-1}$ larger in the
blue and extending $\sim$5\,kpc out from the nucleus.  These emission
lines, as well as other Wolf-Rayet features observed, could be caused
by supernova winds and turbulent motions associated with the merger
\citep{Johansson1988}.  The northern galaxy on the other hand, has
narrower emission lines and weaker starburst activity.

\begin{figure*}[!ht]
\vspace{5mm} \hspace{-3.5mm} \resizebox{\hsize}{!}{\includegraphics{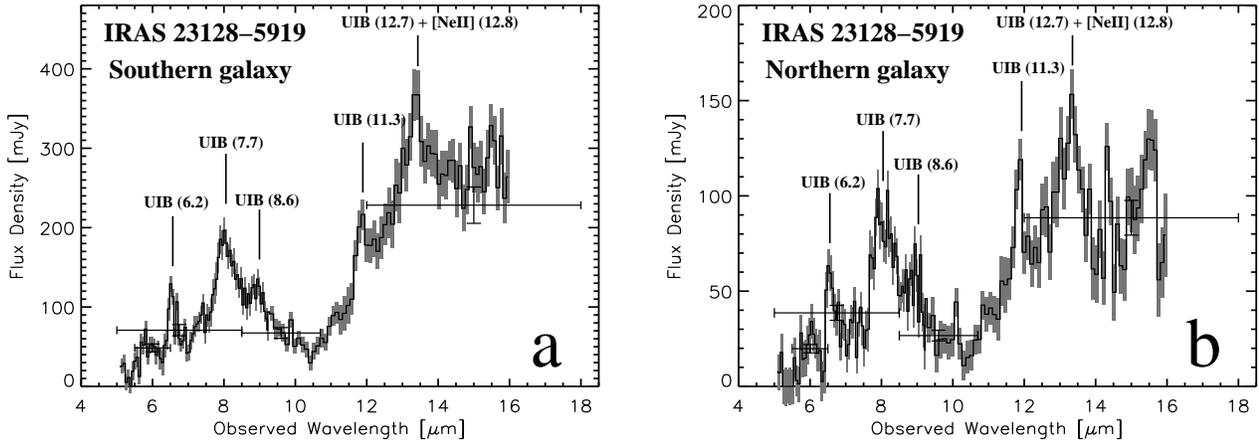}}
\caption{ a) The ISOCAM spectrum of the southern galaxy of IRAS\,23128-5919.
The flux densities of the 4 ISOCAM broad band filters are superimposed
on the CVF spectrum using horizontal lines, the width of which denotes
as usual the filter bandpass. The photometric uncertainties of the CVF
are indicated by the hashed zones. b) The ISOCAM spectrum of the
northern galaxy of IRAS\,23128-5919 using the same convention. All
prominent UIB features redshifted to the distance of the galaxy are
marked.}
\label{ir23128_spec}
\end{figure*}

\begin{figure*}[!ht]
\vspace{5mm} \hspace{-3.5mm} \resizebox{\hsize}{!}{\includegraphics{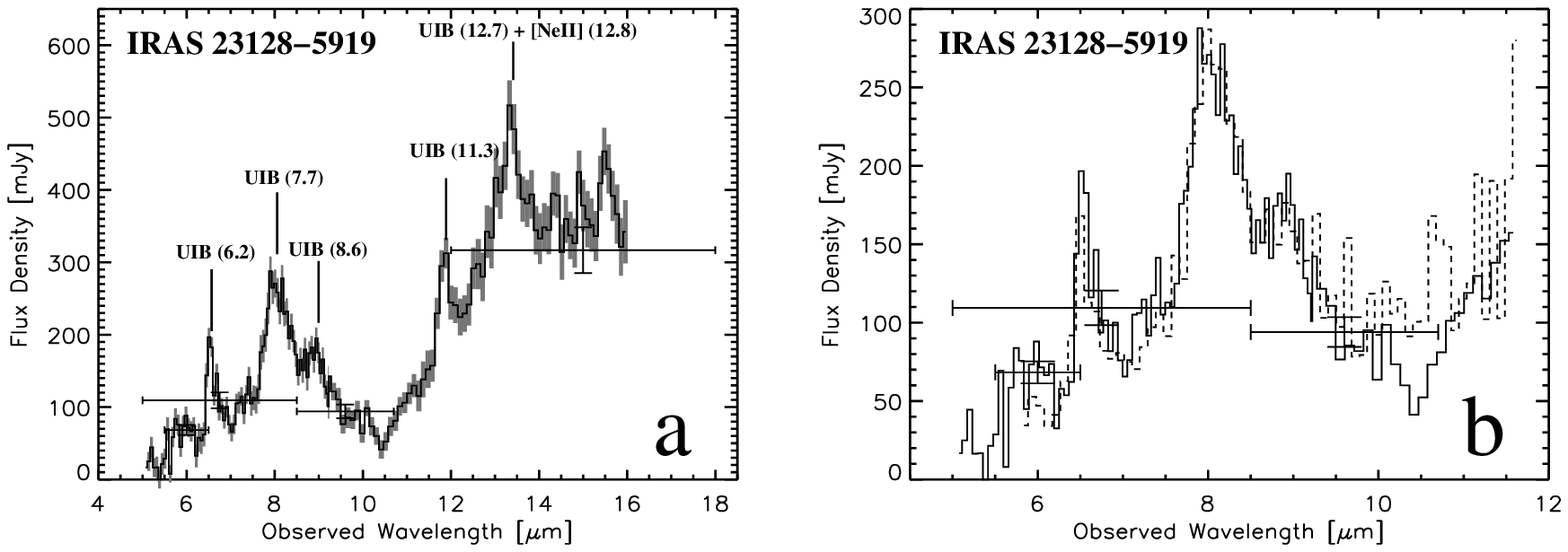}}
\caption{ a) The integrated ISOCAM MIR spectrum of IRAS\,23128-5919 using
the same conventions as in the previous figures. b) The ISOCAM
spectrum of IRAS\,23128-5919 (solid line) superimposed on the
ISOPHOT-S spectrum (dashed line). Our measurements using the ISOCAM
broad band filters are also included.}
\label{ir23128_phot}
\end{figure*}

In Figure~\ref{ir23128_spec}, we present the CVF spectra of each
galaxy along with our flux measurements using the four broad-band
filters. The integrated MIR SED of the whole system is displayed in
Figure~\ref {ir23128_phot}, as well as the ISOPHOT-S spectrum which is
in good agreement with our data. As in the case of IRAS\,19254-7245,
no MIR emission is seen to be associated with the tidal tails down to
our sensitivity limits (see Figure~\ref{ir23128_im}).

We find that approximately 75$\%$ of the MIR flux in IRAS\,23128-5919
originates from the southern galaxy. The spectrum reveals that the
thermal continuum (12--16\,$\mu$m) is higher in the southern galaxy
than that of the north, making the southern galaxy the dominant origin
of the MIR emission. Since the SED of both components displays a
rising spectrum with prominent UIBs and a weak continuum at
5--6\,$\mu$m, we conclude that the MIR emission in this system is
mostly powered by massive star formation. The same conclusion can be
reached using the broad-band filter flux ratios for the two galaxies.
In the northern more quiescent galaxy of the pair, the MIR activity
indicator f$_{15\mu m}$/f$_{6.7\mu m}$ (LW3/LW2) is 2.6, lower than
the value of the southern galaxy ($\sim$\,3.3), while its ratio of
f$_{6.7\mu m}$/f$_{6\mu m}$ is $\sim$\,2.0, higher than that of the
southern galaxy which has an f$_{6.7\mu m}$/f$_{6\mu
m}$$\sim$\,1.5. Following similar reasoning as for the southern
component of the Superantennae, these results can be interpreted as an
increase in the density of \ion{H}{ii} regions of the southern
component, relative to the density of the photo-dissociation
regions. Further comparisons of the properties of this galaxy to
IRAS\,19254-7245 (see Table~\ref{global}) show that its ratio of
L$_{\rm LW3}$/L$_{\rm IR}$ $\sim$\,0.03 is smaller despite is high
L$_{\rm IR}$(L$_{\sun}$)/M$_{\rm H_2}$(M$_{\sun}$) of $\sim$\,70. This
indicates that even though IRAS\,23128-5919 is more efficient in
consuming the molecular gas, its radiation field is not sufficient to
heat the large amount of dust at similarly high temperatures as does
the AGN in the Superantennae. The data presented in Table~\ref{k_ha}
also indicate that the southern galaxy of the pair emits more MIR flux
relatively to its stellar emission (f$_{15\mu m}$/K $\sim$\,284) and
is apparently more obscured by dust (f$_{15\mu m}$/H$\alpha \sim
$\,85).

In conclusion, the more luminous galaxy is clearly undergoing a stronger
star formation phase than its northern companion. The global MIR
characteristics of this system are in agreement with the assertion that a
starburst is the dominant heating mechanism for the dust and no evidence of
an AGN contributing to the ISOCAM wavelength range are present.


\subsection{IRAS\,14348-1447}

\begin{figure*}[!ht]
\vspace{0mm} \hspace{0mm} \resizebox{\hsize}{!}{\includegraphics{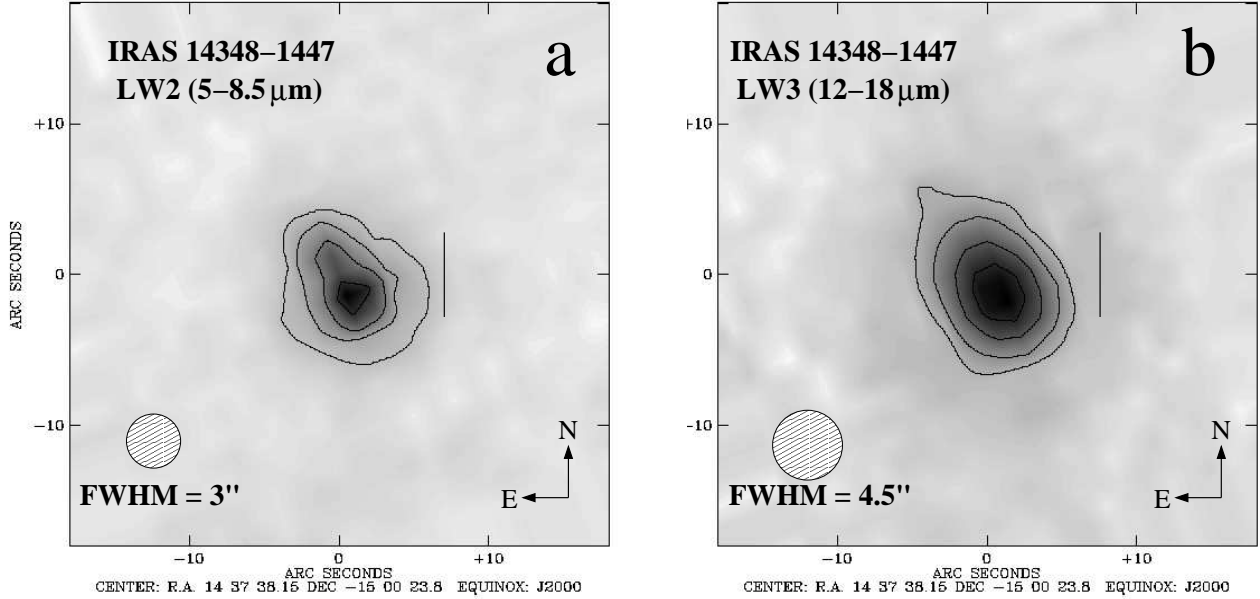}}
\caption{ 
a) Image of IRAS\,14348-1447 observed with ISOCAM at 6.75\,$\mu$m (LW2
filter). The contours are 5, 10, 20, 40 and 80\,$\sigma$
($\sigma$=0.048\,mJy\,pixel$^{-1}$). b) Image of IRAS\,23128-5919
observed with ISOCAM at 15\,$\mu$m (LW3 filter).  The contours are 5,
10, 20 and 30\,$\sigma$ ($\sigma$=0.073\,mJy\,pixel$^{-1}$). The two
vertical bars correspond to 10\,kpc.}
\label{ir14348_im}
\end{figure*}

IRAS\,14348-1447 is the most distant object in the IRAS Bright Galaxy
Sample with a redshift of 0.08 \citep{Soifer1987}. This system, shown
in Figure~\ref{ir14348_im}, consists of two galaxies separated by a
projected distance of 6\,kpc (4\arcsec) with a tail extending to more
than 10\,kpc away from the northern nucleus
\citep{Melnick1990}. Strong H$_2$ emission, mainly triggered by shocks
in molecular clouds, has been detected
\citep{Geballe1988,Nakajima1991}. The presence of large quantities of
shocked molecular hydrogen is consistent with the detection of
6$\times10^{10}$ M{$_{\sun}$} of molecular gas in this system which
makes it the most H$_{2}$-rich in the ultraluminous galaxy sample
\citep{Sanders1991}. The large quantities of cold dust, inferred using
the usual gas to dust conversion, lead us to believe that the
reddening seen in both galaxies is a consequence of strong absorption
and not due to an intrinsically old stellar population
\citep{Carico1990a}.

Based on near-infrared spectroscopic observations in Pa$\alpha$ and
H$_{2}$ lines, the nucleus of the southern galaxy has been classified
as a Seyfert 1.5 and the northern one as a Seyfert 2
\citep{Nakajima1991}, while their optical line features are similar to
those of LINERs \citep{Veilleux1995} or Seyfert 2 galaxies
\citep{Sanders1988}.

Due to its relatively weak MIR emission this source was only observed
with the two ISOCAM broad band filters LW2 and LW3
(Table~\ref{param}). As in the other galaxies in this sample, MIR
emission is detected only from the circumnuclear regions. We estimate
that $\sim$\,75$\%$ of the MIR flux seen in both filters originates
from the southern galaxy, which is also the more active one in the
optical. Interestingly, this roughly scales with the fraction of the
CO emission from the two components \citep{Evans2000}.  The southern
galaxy exhibits the higher hot dust component traced by 15$\mu$m (LW3)
relative to the UIB emission at 7$\mu$m (LW2). Using the LW3/LW2 ratio
to trace the MIR activity in this system we find that f$_{15\mu
m}$/f$_{6.7\mu m}$$\sim$\,3.4 in the southern galaxy and f$_{15\mu
m}$/f$_{6.7\mu m}$ $\sim$\,2.0 in the northern one. Since we only have
one MIR color, we can not comment on the MIR contribution the
AGN. Nevertheless, the low integrated L$_{\rm LW3}$/L$_{\rm IR}$ of
IRAS\,14348-1447 ($\sim$\,0.02)would be consistent with a negligible
AGN contribution in the MIR (Table~\ref{global}) while the high dust
obscuration suggested by the increased f$_{15\mu
m}$/H$\alpha$$\sim$\,333 is consistent with its large molecular gas
content \citep{Mirabel1990}.

\begin{figure}[!ht]
\vspace{0mm} \hspace{0mm} \resizebox{\hsize}{!}{\includegraphics{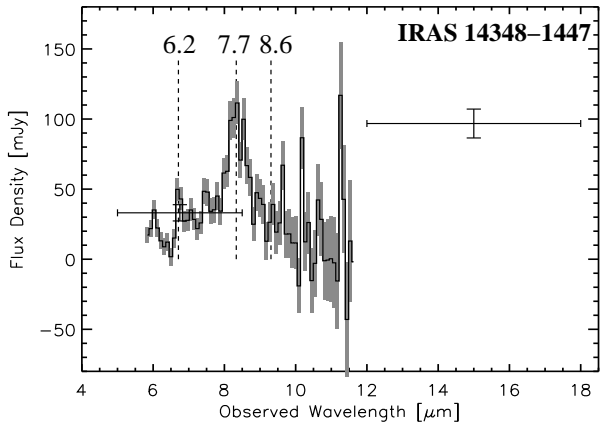}}
\caption{The ISOPHOT-S spectrum of IRAS\,14348-1447 including the
measurements of the two ISOCAM filters. The uncertainties of the ISOPHOT-S
spectrum are indicated by the hashed zones.}
\label{ir14348_phot}
\end{figure}

Evidence that the starburst activity is the main heating mechanism can also
be seen in Figure~\ref{ir14348_phot} using the MIR spectrum of the whole
system obtained with ISOPHOT. This spectrum reveals strong UIBs (f$_{6.7\mu
m}$/f$_{6\mu m}$$\sim$\,2.2, see Table~\ref{phot}) likely caused by a
starburst with only a weak contamination by an AGN \citep[to the 25$\%$
level, see ][]{Genzel1998,Lutz1998}.

\section{Discussion and concluding remarks}

\label{sec:discuss}

A wealth of observational data available has shown that ULIRGs have
high concentrations of gas and dust in their nuclei, sufficient to
account for most of their observed infrared luminosity \citep[see][
for a review]{Sanders1996}. Whether the energy source of ULIRGs is a
dust enshrouded AGN or a starburst still remains an open
issue. However, recent indirect evidence is beginning to favour the
existence of bright extremely red point-like sources in the nuclear
regions of ULIRGs. More specifically near-infrared observations of
luminous infrared galaxies have shown that their flux at 2.2\,$\mu$m
is more concentrated towards the center than at 1.3\,$\mu$m
\citep{Carico1990b,Scoville2000}. Furthermore, recent high resolution MIR observations
using Keck of a sample of ULIRGs reveal compact sub-arcsecond sources
(with linear scales of $\sim$\,100--300~pc) which contain 30\% to
100\% of the observed MIR energy of these galaxies
\citep{Soifer2000}. This contrasts with the LIRGs
($10^{11}$\,L$_{\sun}$ $\leq$ L$_{\rm IR} \leq 10^{12}$\,L$_{\sun}$),
in which the infrared energy seems to be generated over somewhat
larger scales \citep[$\sim$\,100\,pc\---1\,kpc,][]{Soifer2001} and
sometimes can be found in extra-nuclear regions associated with the
physical interaction of merging pairs of galaxies. Furthermore, there
are galaxies such as VV\,114 where it has even been found that a
substantial fraction of the MIR flux originates from an extended
component of hot dust emission spread over several kpc scales
\citep{Soifer2001,LeFloch2002}. ULIRGs are thus not simply a scaled-up
version of LIRGs and require further dynamical compression of the
molecular gas responsible for the IR luminosity within very compact
regions. A plausible mechanism would be one where the shocks and tidal
forces of the interaction first lead to star formation over galactic
scales, leading to IR luminosities up to a few
10$^{11}$\,L$_{\sun}$. Subsequently, gravitational instabilities and
the formation of a bar, strip the gas of its angular momentum,
funneling large quantities towards the nuclear regions of galaxies,
which can feed circumnuclear starbursts or AGNs and trigger the
ultraluminous phase in the infrared \citep{Combes2001}.

Even though the above scenario is appealing, given the high extinction
in the nuclei of ULIRGs, the limited atmospheric transmission in the
MIR windows, and the limited sensitivity of ground-based instruments,
questions related to the direct probing of the nuclear activity such
as ``does all MIR emission from those systems originate from the
nuclei?'' and if not ``what are the spectral properties of any
extended component?''  still remain unanswered.

This is where the superb sensitivity of space instruments, such as
ISO, is essential.  We have found that in the ULIRGs studied here
\emph{more than $\sim $\,95\% of the MIR emission seen by IRAS is
confined within a few arcsecs of their central region}.  Obviously the
relatively large pixel size of the ISOCAM detector places limitations
in interpreting these findings. However, deconvolution tests of the
central point source in each galaxy suggest that the corresponding
nuclei are resolved and the physical diameter of the emitting region
is contained within 1 to 2\,kpc. Moreover, with the exception of the
Superantennae where the MIR spectrum is dominated by the emission
arising from the AGN of the southern galaxy, the bulk of the IR
luminosity of IRAS\,23128-5919 and IRAS\,14348-1447 is powered by
massive star formation. The fact that starbursts can dominate the MIR
emission in galaxies with IR luminosities as high
$\sim$10$^{12}$\,L$_{\sun}$ had already been demonstrated in other
ISOCAM-CVF \citep{Tran2001} and ISO-SWS \citep{Genzel1998}
observations of ULIRGs, and is supported by our results.  Given that
an active nucleus appears to be always present in the most energetic
objects of the local Universe \citep{Lutz1998}, our MIR data favor a
luminosity threshold for the transition between starburst- and
AGN-dominated galaxies which is higher than the IR luminosity of the
galaxies in our sample. This is in agreement with the results of
\citet{Tran2001} who proposed that this transition takes place at
L$_{\rm IR}\sim$10$^{12.5}$\,L{$_{\sun}$} and also found individual
starbursts up to 10$^{12.65}$\,L{$_{\sun}$}. Our data also indicate is
that such starbursts can be confined to the very central nuclear
regions which may have important consequences in the probing how the
instabilities fuel the inner regions of galaxies
\citep[e.g.][]{Combes2001}, as well as determining the nature of high
redshift dusty sources \citep[e.g.][]{Ivison2000}.

Another striking feature revealed in our observations is that in all
three cases one galaxy seems to dominate the MIR energy output of the
system by more than 75\%. Could this be a record of the initial
distribution of the amount of molecular gas available in each merging
progenitor or could this suggest that in the later stages of
interaction, the gas finally merges towards \textit{one} component? If
the latter were true one would expect that a sufficiently large
quantity of gas could trigger and fuel both circumnuclear star forming
activity and AGN-type activity at the core of a single object. This is
evident in the southern galaxy of IRAS\,19254-7245 which harbors an
active nucleus as well as numerous massive star forming regions. As we
mentioned in the introduction though the presence of a Seyfert nucleus
is correlated with a MIR flux increase relative to the FIR luminosity
of the entire galaxy, which is what one can actually derive from our
observations when we compare the Superantennae with
IRAS\,14348-1447. IRAS\,14348-1447 has indeed a much higher total IR
luminosity despite its MIR flux being lower than that of the southern
source of IRAS\,19254-7245.  Furthermore, using the f$_{15\mu
m}$/H$\alpha$ and f$_{15\mu m}$/K ratios as probes of dust absorption
and hot dust emission normalized to the mass of the galaxy, we find
that in each interacting system it is always the most active galaxy of
the system that exhibits the higher ratios. In each system, the most
luminous galaxy contains a larger amount of molecular gas leading to
the triggering/feeding of the starburst activity and/or an active
nucleus.

Finally, we wish to stress once more that because of the limited
spatial resolution in studying such distant sources, the diagnostics
we have used in this paper address only the integrated MIR emission of
each galaxy. Our difficulties to identify whether an active nucleus is
solely responsible for the increase in the MIR luminosity relative to
the FIR emission will not be resolved unless we can either clearly map
the extent of the emitting region or obtain MIR spectra using very
narrow slits. The upcoming launch of SIRTF which, despite the fact it
has comparable spatial resolution to ISO, is equipped with a new
generation of detectors of smaller pixel size, and in particular the
use of its infrared spectrograph will help us improve upon our current
results and provide conclusive answers to the issues which still
remain unresolved to date.

\begin{acknowledgements}
We wish to thank P.-A Duc for providing his K band images of
IRAS\,19254-7245 and IRAS\,23128-5919 as well as D. Rigopoulou for
providing the ISOPHOT data. We greatly appreciated the comments of the
referee which helped us improve the manuscript. VC would like to
acknowledge the partial support of JPL contract 960803.

\end{acknowledgements}


\begin{thebibliography}{}

\bibitem[Barvainis(1987)]{Barvainis1987}   
	Barvainis, R. 1987, \apj, 320, 537

\bibitem[Bergvall \& Johansson(1985)]{Bergvall1985}   
	Bergvall, N., \& Johansson, L. 1985, \aap, 149, 475

\bibitem[Borne et~al.(1999)]{Borne1999}   
	Borne, K.~D., Bushouse, H., Colina, L., et al. 1999, \apss,
	266, 137

\bibitem[Boselli et~al.(1997)]{Boselli1997}   
	Boselli, A., Lequeux, J., Contursi, A., et al. 1997, \aap,
	324, L13

\bibitem[Bryant \& Scoville(1999)]{Bryant1999}   
	Bryant, P. M., \& Scoville, N. Z. 1999, \aj, 117, 2632

\bibitem[Carico et~al.(1990a)]{Carico1990a}   
	Carico, D.~P., Graham, J.~R., Matthews, K., et al. 1990a,
	\apjl, 349, L39

\bibitem[Carico et~al.(1990b)]{Carico1990b}   
	Carico, D.~P., Sanders, D.~B., Soifer, B.~T., Matthews, K., \&
	Neugebauer, G. 1990b, \aj, 100, 70

\bibitem[Cesarsky et~al.(1996a)]{CesarskyC1996}   
	Cesarsky, C.~J., Abergel, A., Agnese, P., et al. 1996a, \aap,
	315, L32

\bibitem[Cesarsky et~al.(1996b)]{CesarskyD1996b}   
	Cesarsky, D., Lequeux, J., Abergel, A., et al. 1996b, \aap,
	315, L305

\bibitem[Charmandaris et~al.(1999a)]{Charmandaris1999a}   
	Charmandaris, V., Laurent, O., Mirabel, I.~F., et al. 1999a,
	\aap, 341, 69

\bibitem[Charmandaris et~al.(1999b)]{Charmandaris1999b}   
	Charmandaris, V., Laurent, O., Mirabel, I.~F., et al. 1999b,
	\apss, 266, 99

\bibitem[Charmandaris et~al.(2002)]{Charmandaris2002}   
	Charmandaris, V., Le Floc'h, E., Laurent, O., et al. 2002,
	\apj, (in preparation)

\bibitem[Clavel et~al.(2000)]{Clavel2000}   
	Clavel, J., Schulz, B., Altieri, B., et al. 2000, \aap, 357,
	839

\bibitem[Colina et~al.(1991)]{Colina1991}   
	Colina, L., Lipari, S., \& Macchetto, F. 1991, \apj, 379, 113

\bibitem[Combes(2001)]{Combes2001}   
	Combes, F. 2001, Fueling the AGN. In Lectures on the
	Starburst-AGN Connection, INAOE, ed. D. Kunth, I. Aretxaga
	[astro-ph/0010570]

\bibitem[Condon et~al.(1991)]{Condon1991}   
	Condon, J.~J., Huang, Z.-P., Yin, Q.~F., \& Thuan, T.~X. 1991,
	\apj, 378, 65

\bibitem[Coulais \& Abergel(2000)]{Coulais2000}   
	Coulais, A., \& Abergel, A. 2000, \aaps, 141, 533

\bibitem[de Grijp et~al.(1985)]{deGrijp1985}   
	de Grijp, M.~H.~K., Miley, G.~K., Lub, J., \& de Jong,
	T. 1985, \nat, 314, 240

\bibitem[Dale et al.(2001)]{Dale2001}   
	Dale, D. A., Helou, G., Contursi, A., Silbermann, N. A., \&
	Kolhatkar, S. 2001, \apj, 549, 215

\bibitem[Duc \& Mirabel(1997a)]{Duc1997a}   
	Duc, P.~A., \& Mirabel, I.~F.  1997a, The Messenger, 89, 14

\bibitem[Duc et~al.(1997b)]{Duc1997b}  
	Duc, P.~A., Mirabel, I.~F., \& Maza, J. 1997b, \aaps, 124, 533

\bibitem[Dudley(1999)]{Dudley1999}   
	Dudley, C.~C. 1999, \mnras, 307, 553

\bibitem[Evans et~al.(2000)]{Evans2000}   
	Evans, A. S., Surace, J. A., \& Mazzarella, J. M. 2000, \apj,
	529, L88

\bibitem[Geballe(1988)]{Geballe1988}   
	Geballe, T.~R. 1988, \mnras, 234, 1P

\bibitem[Genzel et~al.(1998)]{Genzel1998}   
	Genzel, R., Lutz, D., Sturm, E., et al. 1998, \apj, 498, 579

\bibitem[Helou et al.(2001)]{Helou2001}   
	Helou, G., Malhotra, S., Hollenbach, D. J., Dale, D. A., \&
	Contursi, A. 2001, \apj, 549, 215

\bibitem[Houck et~al.(1984)]{Houck1984}   
	Houck, J.~R., Soifer, B.~T., Neugebauer, G., et al. 1984,
	\apjl, 278, L63

\bibitem[Imanishi \& Dudley(2000)]{Imanishi2000}   
	Imanishi, M., \& Dudley, C.~C. 2000, \apj, 545, 701

\bibitem[Ivison et al.(2000)]{Ivison2000}   
	Ivison R. J., Smail I., Barger A. J., et al. 2000, \mnras,
	315, 209

\bibitem[Johansson \& Bergvall(1988)]{Johansson1988}   
	Johansson, L., \& Bergvall, N. 1988, \aap, 192, 81

\bibitem[Joseph(1999)]{Joseph99}   
	Joseph, R. D. 1999, \apss, 266, 321

\bibitem[Kessler et~al.(1996)]{Kessler1996}   
	Kessler, M.~F., Steinz, J.~A., Anderegg, M.~E., et al. 1996,
	\aap, 315, L27

\bibitem[Laurent(1999a)]{Laurent1999a}   
	Laurent, O. 1999a, Ph.D. Thesis, University of Paris~XI,
	France

\bibitem[Laurent et al.(1999b)]{Laurent1999b}   
	Laurent O., Mirabel I.F., Charmandaris V., et al. 1999b, in
	XIXth Moriond Astrophysics Meeting.  Building the Galaxies:
	From the Primordial Universe to the Present, 79
	[astro-ph/0005377]

\bibitem[Laurent et~al.(2000)]{Laurent2000}   
	Laurent, O., Mirabel, I.~F., Charmandaris, V., et al. 2000,
	\aap, 359, 887

\bibitem[Le Floc'h et~al.(2001)]{LeFloch2001}   
	Le Floc'h, E., Mirabel, I.~F., Laurent, O., et al. 2001, \aap,
	367, 487

\bibitem[Le Floc'h et~al.(2002)]{LeFloch2002}   
	Le Floc'h, E., Charmandaris V., Laurent, O., et al. 2002,
	\aap, (in press) [astro-ph/0205401]

\bibitem[Leger et~al.(1989)]{Leger1989}   
	Leger, A., D'Hendecourt, L., Boissel, P., \& Desert,
	F.~X. 1989, \aap, 213, 351

\bibitem[Lutz et~al.(1998)]{Lutz1998}   	
	Lutz, D., Spoon, H. W.~W., Rigopoulou, D., Moorwood, A. F.~M.,
	\& Genzel, R. 1998, \apjl, 505, L103

\bibitem[Madden et~al.(1997)]{Madden1997}   
	Madden, S.~C., Vigroux, L., \& Sauvage, M. 1997, in
	Extragalactic Astronomy in the Infrared, ed. G.~A.  Mamon,
	T.~X. Thuan, \& J. Tran Thanh Van, 229

\bibitem[Mathis(1990)]{Mathis1990}   
	Mathis, J.~S. 1990, \araa, 28, 37

\bibitem[Melnick \& Mirabel(1990)]{Melnick1990}   
	Melnick, J., \& Mirabel, I.~F. 1990, \aap, 231, L19

\bibitem[Mihos \& Bothun(1998)]{Mihos1998}   
	Mihos, J.~C., \& Bothun, G.~D.  1998, \apj, 500, 619

\bibitem[Mirabel et~al.(1990)]{Mirabel1990}   
	Mirabel, I.~F., Booth, R.~S., Johansson, L. E.~B., Garay, G.,
	\& Sanders, D.~B. 1990, \aap, 236, 327

\bibitem[Mirabel et~al.(1991)]{Mirabel1991}   
	Mirabel, I.~F., Lutz, D., \& Maza, J. 1991, \aap, 243, 367

\bibitem[Mirabel et~al.(1998)]{Mirabel1998}   
	Mirabel, I.~F., Vigroux, L., Charmandaris, V., et al. 1998,
	\aap, 333, L1

\bibitem[Nakajima et~al.(1991)]{Nakajima1991}   
	Nakajima, T., Kawara, K., Nishida, M., \& Gregory, B. 1991,
	\apj, 373, 452

\bibitem[Pier \& Krolik(1992)]{Pier1992}   
	Pier, E.~A., \& Krolik, J.~H.  1992, \apj, 401, 99

\bibitem[Richstone et~al.(1998)]{Richstone1998}   
	Richstone, D., Ajhar, E.~A., Bender, R., et al. 1998, \nat,
	395, A14

\bibitem[Rigopoulou et~al.(1999)]{Rigopoulou1999}   
	Rigopoulou, D., Spoon, H. W.~W., Genzel, R., et al. 1999, \aj,
	118, 2625

\bibitem[Roche et~al.(1991)]{Roche1991}   
	Roche, P.~F., Aitken, D.~K., Smith, C.~H., \& Ward,
	M.~J. 1991, \mnras, 248, 606

\bibitem[Roussel et~al.(2001)]{Roussel2001}   
	Roussel, H., Sauvage, M., Vigroux, L., \& Bosma, A. 2001,
	\aap, 372, 427

\bibitem[Sanders \& Mirabel(1985)]{Sanders1985}   
	Sanders, D.~B., \& Mirabel, I.~F. 1985, \apj, 298, 31

\bibitem[Sanders et al.(1986)]{Sanders1986}   
	Sanders, D. B., Scoville, N. Z., Young, J. S., et al.  \apj,
	305 45L

\bibitem[Sanders et~al.(1988)]{Sanders1988}   
	Sanders, D.~B., Soifer, B.~T., Elias, J.~H., et al. 1988,
	\apj, 325, 74

\bibitem[Sanders et~al.(1991)]{Sanders1991}   
	Sanders, D.~B., Scoville, N.~Z., \& Soifer, B.~T. 1991, \apj,
	370, 158

\bibitem[Sanders \& Mirabel(1996)]{Sanders1996}   
	Sanders, D.~B., \& Mirabel, I.~F. 1996, \araa, 34, 749

\bibitem[Sanders(1999)]{Sanders99}   
	Sanders, D. B. 1999, \apss, 266, 331

\bibitem[Sauvage et~al.(1996)]{Sauvage1996}   
	Sauvage, M., Blommaert, J., Boulanger, F., et al. 1996, \aap,
	315, L89

\bibitem[Scoville et~al.(2000)]{Scoville2000}   
	Scoville, N.~Z., Evans, A.~S., Thompson, R., et al. 2000, \aj,
	119, 991

\bibitem[Soifer et~al.(1987)]{Soifer1987}   
	Soifer, B.~T., Sanders, D.~B., Madore, B.~F., et al. 1987,
	\apj, 320, 238

\bibitem[Soifer et~al.(1989)]{Soifer1989}   
	Soifer, B.~T., Boehmer, L., Neugebauer, G., \& Sanders,
	D.~B. 1989, \aj, 98, 766

\bibitem[Soifer et~al.(2000)]{Soifer2000}   
	Soifer, B.~T., Neugebauer, G., Matthews, K., et al. 2000, \aj,
	119, 509

\bibitem[Soifer et~al.(2001)]{Soifer2001}   
	Soifer, B.~T., Neugebauer, G., Matthews, K., et al. 2001, \aj,
	122, 1213

\bibitem[Starck et~al.(1997)]{Starck1997}   
	Starck, J.~L., Siebenmorgen, R., \& Gredel, R. 1997, \apj,
	482, 1011

\bibitem[Starck et~al.(1999)]{Starck1999}   
	Starck, J.~L., Abergel, A., Aussel, H., et al. 1999, \aaps,
	134, 135

\bibitem[Wild et al.(1992)]{Wild1992}
	Wild, W., Harris, A. I., Eckart, A., et al. 1992, \aap, 265,
	447

\bibitem[Tran et~al.(2001)]{Tran2001}   
	Tran, Q.~D., Lutz, D., Genzel, R., et al. 2001, \apj, 552, 527

\bibitem[Vanzi et al.(2002)]{Vanzi2002}   
	Vanzi, L., Bagnulo, S., Le Floc'h, E., et al. 2002, \aap, 386,
	464

\bibitem[Veilleux et~al.(1995)]{Veilleux1995}   
	Veilleux, S., Kim, D.-C., Sanders, D.~B., Mazzarella, J.~M.,
	\& Soifer, B.~T. 1995, \apjs, 98, 171

\bibitem[Verstraete et~al.(1996)]{Verstraete1996}   
	Verstraete, L., Puget, J.-L., Falgarone, E., et al. 1996,
	\aap, 315, L337

\bibitem[Vigroux et~al.(1999)]{Vigroux1999}   
	Vigroux, L., Charmandaris, V., Gallais, P., et al. 1999, in
	The Universe as Seen by ISO, ESA SP-427, 805

\end{thebibliography}
\end{document}